\documentclass[aps,manuscript,reprint,noshowpacs,noshowkeys,prd,balancelastpage,nofootinbib]{revtex4}
\usepackage[utf8]{inputenc}
\usepackage{color}
\usepackage{amsmath}
\usepackage{amssymb}
\usepackage[unicode=true,
 bookmarks=false,
 breaklinks=false,pdfborder={0 0 1},backref=section,colorlinks=true]
 {hyperref}
\hypersetup{
 linkcolor=blue,urlcolor=blue,citecolor=red}

\makeatletter
\@ifundefined{textcolor}{}
{%
 \definecolor{BLACK}{gray}{0}
 \definecolor{WHITE}{gray}{1}
 \definecolor{RED}{rgb}{1,0,0}
 \definecolor{GREEN}{rgb}{0,1,0}
 \definecolor{BLUE}{rgb}{0,0,1}
 \definecolor{CYAN}{cmyk}{1,0,0,0}
 \definecolor{MAGENTA}{cmyk}{0,1,0,0}
 \definecolor{YELLOW}{cmyk}{0,0,1,0}
}

\usepackage{amsfonts}
\usepackage{mdframed}
\usepackage{footnote}
\usepackage{float}
\usepackage[font=footnotesize,it]{caption}
\usepackage{tikz}
\usepackage{tikz-3dplot}

\makeatother

\begin{document}
\title{Generalized probability and current densities: A field theory approach}
\author{M. Izadparast }
\email{masoumeh.izadparast@emu.edu.tr}

\author{S. Habib Mazharimousavi}
\email{habib.mazhari@emu.edu.tr}

\affiliation{Department of Physics, Faculty of Arts and Sciences, Eastern Mediterranean
University, Famagusta, North Cyprus via Mersin 10, Turkey}
\date{\today }
\begin{abstract}
We introduce a generalized Lagrangian density - involving a non-Hermitian
kinetic term - for a quantum particle with the generalized momentum
operator. Upon variation of the Lagrangian, we obtain the corresponding
Schrödinger equation. The extended probability and particle current
densities are found which satisfy the continuity equation.
\end{abstract}
\maketitle

\section{Introduction}

The idea of the generalized momentum operator has been extended in
our earlier proposal \cite{M.H1,M.H2}, which is subjected to the
non-relativistic and non-Hermitian quantum mechanics. The thought
of the non-Hermitian Hamiltonian was discussed initially in Ref. \cite{Bessis,Caliceti}.
Yet, Bender and Boettcher have initiated $\mathcal{PT}$-symmetric
quantum physics by employing a harmonic-oscillator-like $\mathcal{PT}$-symmetric
potential as a toy model \cite{Bender1}. They have shown that even
if the Hamiltonian is non-Hermitian the eigenvalues can be real, upon
which, a $\mathcal{PT}$-symmetric operator is defined. The $\mathcal{PT}$
operator is composed of the parity and the time reversal operators,
namely i. e., $\mathcal{P}x\mathcal{P}=-x$ and $\mathcal{T}i\mathcal{T}=-i$
\cite{Bender2}. For a $\mathcal{PT}$-symmetric Hamiltonian - a less
general non-Hermitian Hamiltonian - the interaction potential is considered
to be complex. Hence, the unitarity for the unbroken $\mathcal{PT}$-symmetry
(under which the energy spectrum of the system is real) in a quantum
mechanical system has to be conserved, which is one of the prominent
principles of quantum mechanics. Accordingly, in Ref. \cite{Bender3}
the infinitesimal probability density is considered and a method is
introduced to find the path integral in the complex plane $\mathbb{C}$.
Earlier in Ref. \cite{Bagchi}, Bagchi et. al. discussed the generalized
continuity equation in accordance with the Schrödinger equation and
its $\mathcal{PT}$-symmetric conjugate. Furthermore, the modified
normalization constant is obtained on the real $x$-axis. In the approach
of the field theory upon applying the principle of stationary action,
one finds the equations of motion of a particle in the classical or
quantum mechanical system. There exist various studies in the literature
on the generalization of Lagrangian density which well-describe different
physical phenomenologies. According to Ref. \cite{El-Nabulsi}, non-standard
power-law Lagrangians are discussed following the generalization of
the fractional calculus of variation. In this framework, the dissipative
classical and quantum dynamical systems are not well-defined with
the standard Lagrangian, whilst the fractional action-like scheme
represents a successful method to deal with such applications in physics.
Nobre et. al. in Ref. \cite{Nobre1} expressed the non-linear Schrödinger
equation upon imposing the non-linearity into the kinetic energy.
Furthermore, they found the corresponding Schrödinger equation by
applying the variation of Lagrangian density, also, they identified
the necessity of a new second field to generate the Schrödinger equation.
The authors in Ref. \cite{Nobre2,Rego1,Plastino} have expressed a
set of non-Hermitian $\mathcal{PT}$-symmetric Schrödinger equations
corresponding to the position-dependent mass in classical field theory.
The probability density has been extracted based on the definition
of the continuity equation for the linear Schrödinger equation in
Ref. \cite{Nobre1,Nobre2,Plastino}. Rego-Monteiro et. al., in Ref.
\cite{Rego2}, have analyzed a general family of non-self-adjoint
Hamiltonians corresponding to the position-dependent mass which describe
some phenomenological aspects of the condensed matter. Also in Ref.
\cite{Spourdalakis}, a Lagrangian, which is constructed by definition
of two complex fields, has been introduced. With the variational approach,
the generalized continuity equations have been derived in the Hermitian
and non-Hermitian quantum mechanical systems. 

Here in this paper, we introduce a generalized Lagrangian density
appointed to the generalized momentum operator \cite{M.H1,M.H2}.
We also study the physical aspects of the Lagrangian density including
the corresponding Schrödinger equation stress-energy tensor and the
probability density.

The present paper is organized as follows. In Sec. II we introduce
the generalized Lagrangian density which corresponds to the extended
definition of the generalized momentum operator. Besides that, by
applying the principle of stationary action through the Euler-Lagrange
equation, we obtain the generalized Hamiltonian and the corresponding
Schrödinger equation. In Sec. III the continuity equation is obtained
under the concept of the generalized probability and current densities.
Finally, we summarize our paper in the Conclusion.

\section{The Lagrangian Density }

Upon recalling the generalized momentum operator presented in Ref.
\cite{M.H1,M.H2}, i.e 
\begin{equation}
\hat{p}=-i\hbar\left(\left(1+\mu\right)\partial_{x}+\frac{\mu^{\prime}}{2}\right),\label{generalized-momentum}
\end{equation}
 where the auxiliary function $\mu\left(x\right)$ is a $\mathcal{PT}$-symmetric
function of $x$, we introduce the generalized Lagrangian density
to be 
\begin{align}
\mathcal{\mathcal{L}=}\frac{i\hbar}{2}\phi\dot{\psi}-\frac{\hbar^{2}}{4m}\left[\left(1+\mu\right)^{2}\phi^{\prime}\psi^{\prime}+\left(1+\mu\right)\frac{\mu^{\prime}}{2}\left(\phi\psi\right)^{\prime}++\frac{\mu^{\prime2}}{4}\phi\psi\right]-\frac{1}{2}V\left(x,t\right)\phi\psi-\nonumber \\
\frac{i\hbar}{2}\phi^{*}\dot{\psi^{*}}-\frac{\hbar^{2}}{4m}\left[\left(1+\mu^{*}\right)^{2}\phi^{*\prime}\psi^{*\prime}+\left(1+\mu^{*}\right)\frac{\mu^{*\prime}}{2}\left(\phi^{*}\psi^{*}\right)^{\prime}+\frac{\mu^{*\prime2}}{4}\phi^{*}\psi^{*}\right]-\frac{1}{2}V^{*}\left(x,t\right)\phi^{*}\psi^{*}.\nonumber \\
\label{Lagrangian-generalized}
\end{align}
Herein, a dot and a prime stand for the derivative with respect to
$t$ and $x$, respectively, and {*} implies the complex conjugate.
We note that, following \cite{Nobre1,Nobre2,Plastino,Rego1,Rego2,Spourdalakis},
the Lagrangian density \eqref{Lagrangian-generalized} is constructed
using two field functions $\phi$ and $\psi$ and their complex conjugate,
such that by choosing $\phi=\psi^{*}$ and $\mu=0$ with a real interaction
potential $V\left(x,t\right)$, it reduces to the standard Schrödinger
Lagrangian density which is given by 
\begin{equation}
\mathcal{\mathcal{L}=}\frac{i\hbar}{2}\left(\psi^{*}\dot{\psi}-\psi\dot{\psi^{*}}\right)-\frac{\hbar^{2}}{2m}\psi^{*\prime}\psi^{\prime}-V\left(x,t\right)\psi\psi^{*}.\label{Schrodinger-Lagrangian}
\end{equation}
 Referring to \eqref{Lagrangian-generalized} the potential is assumed
to be $\mathcal{PT}$-symmetric i. e. $V\left(x,t\right)=V^{*}\left(-x,t\right)$.
Variation of the action with respect to $\phi$ and $\psi$ yields
the field equations which are expressed as

\begin{equation}
i\hbar\dot{\psi}=-\frac{\hbar^{2}}{2m}\left(\left(1+\mu\right)^{2}\partial_{x}^{2}+2\left(1+\mu\right)\mu^{\prime}\partial_{x}+\frac{\mu^{\prime\prime}}{2}\left(1+\mu\right)+\frac{\mu^{\prime2}}{4}\right)\psi+V\left(x,t\right)\psi\label{Schrodinger-psi}
\end{equation}
and
\begin{equation}
-i\hbar\dot{\phi}=-\frac{\hbar^{2}}{2m}\left(\left(1+\mu\right)^{2}\partial_{x}^{2}+2\left(1+\mu\right)\mu^{\prime}\partial_{x}+\frac{\mu^{\prime\prime}}{2}\left(1+\mu\right)+\frac{\mu^{\prime2}}{4}\right)\phi+V\left(x,t\right)\phi,\label{Schrodinger-phi}
\end{equation}
respectively. Eq. \eqref{Schrodinger-psi} is the generalized Schrödinger
equation studied in Ref. \cite{M.H1,M.H2} while \eqref{Schrodinger-phi}
is the $\mathcal{PT}$ conjugate equation of \eqref{Schrodinger-psi}
provided $\phi=\psi^{\mathcal{PT}}$. Therefore, the field equations
\eqref{Schrodinger-psi} and \eqref{Schrodinger-phi}can be summarized
as
\begin{equation}
\begin{cases}
i\hbar\dot{\psi}=\hat{H}\psi\\
-i\hbar\dot{\phi}=\hat{H}\phi
\end{cases},\label{Conjugate}
\end{equation}
where the Hamiltonian operator $\hat{H}$ is defined to be 
\begin{equation}
\hat{H}=\hat{H}^{\mathcal{PT}}=-\frac{\hbar^{2}}{2m}\left(\left(1+\mu\right)^{2}\partial_{x}^{2}+2\left(1+\mu\right)\mu^{\prime}\partial_{x}+\frac{\mu^{\prime\prime}}{2}\left(1+\mu\right)+\frac{\mu^{\prime2}}{4}\right).\label{Original-Ham}
\end{equation}
Moreover, the field equations corresponding to $\psi^{*}$ and $\phi^{*}$
are obtained to be 
\begin{equation}
-i\hbar\dot{\psi^{*}}=-\frac{\hbar^{2}}{2m}\left(\left(1+\mu^{*}\right)^{2}\partial_{x}^{2}+2\left(1+\mu^{*}\right)\mu^{\prime*}\partial_{x}+\frac{\mu^{*\prime\prime}}{2}\left(1+\mu^{*}\right)+\frac{\mu^{*\prime2}}{4}\right)\psi^{*}+V^{*}\left(x,t\right)\psi^{*}\label{Schrodinger-star}
\end{equation}
and
\begin{equation}
i\hbar\dot{\phi^{*}}=-\frac{\hbar^{2}}{2m}\left(\left(1+\mu^{*}\right)^{2}\partial_{x}^{2}+2\left(1+\mu^{*}\right)\mu^{\prime*}\partial_{x}+\frac{\mu^{*\prime\prime}}{2}\left(1+\mu^{*}\right)+\frac{\mu^{*\prime2}}{4}\right)\phi^{*}+V^{*}\left(x,t\right)\phi^{*}.\label{Shrodinger-star2}
\end{equation}
 Next, using the Lagrangian density \eqref{Lagrangian-generalized}
and the definition of Hamiltonian density 
\begin{equation}
\mathcal{H}=\Sigma_{\sigma}\pi_{\sigma}\dot{f_{\sigma}}-\mathcal{L},\label{Hamiltonian-den}
\end{equation}
 where $f_{\sigma}\in\left\{ \psi,\psi^{*},\phi,\phi^{*}\right\} $
and $\pi_{\sigma}$ is the momentum-density conjugate to $f_{\sigma}$
we calculate the explicit form of $\mathcal{H}$. To do so, let's
calculate $\pi_{\sigma}$ as 

\begin{equation}
\begin{cases}
\pi_{\psi}=\frac{\partial\mathcal{L}}{\partial\dot{\psi}}=\frac{i\hbar}{2}\phi, & \pi_{\phi}=\frac{\partial\mathcal{\mathcal{L}}}{\partial\dot{\phi}}=0,\\
\pi_{\psi^{*}}=\frac{\partial\mathcal{L}}{\partial\dot{\psi^{*}}}=\frac{i\hbar}{2}\phi^{*}, & \pi_{\phi^{*}}=\frac{\partial\mathcal{\mathcal{L}}}{\partial\dot{\phi^{*}}}=0.
\end{cases}\label{canonical momenta}
\end{equation}
After the substitution into Eq. \eqref{Hamiltonian-den}, the one
dimensional Hamiltonian density is obtained to be 
\begin{align}
\mathcal{H} & =\frac{\hbar^{2}}{4m}\left[\left(1+\mu\right)^{2}\phi^{\prime}\psi^{\prime}+\frac{\mu^{\prime}\left(1+\mu\right)}{2}\left(\phi\psi\right)^{\prime}+\frac{\mu^{\prime2}}{4}\phi\psi\right]+\frac{1}{2}V\left(x,t\right)\phi\psi+\label{Hamiltonian-density}\\
 & \frac{\hbar^{2}}{4m}\left[\left(1+\mu^{*}\right)^{2}\phi^{*\prime}\psi^{*\prime}+\frac{\mu^{*\prime}\left(1+\mu^{*}\right)}{2}\left(\phi^{*}\psi^{*}\right)^{\prime}+\frac{\mu^{*\prime2}}{4}\phi^{*}\psi^{*}\right]+\frac{1}{2}V^{*}\left(x,t\right)\phi^{*}\psi^{*}.\nonumber 
\end{align}
Having hamiltonian density found, we apply $E=\int\mathcal{H}dx$
to find the energy of the system. Explicitly, one finds 
\begin{align*}
E & =\int\left[\frac{-\hbar^{2}}{8m}\psi\left(\left(1+\mu\right)^{2}\partial_{x}^{2}+2\mu^{\prime}\left(1+\mu\right)\partial_{x}+\frac{\mu^{\prime\prime}}{2}\left(1+\mu\right)+\frac{\mu^{\prime2}}{4}\right)\phi+\psi\frac{1}{4}V\left(x,t\right)\phi\right]dx+\\
 & \int\left[-\frac{\hbar^{2}}{8m}\phi\left(\left(1+\mu\right)^{2}\partial_{x}^{2}+2\mu^{\prime}\left(1+\mu\right)\partial_{x}+\frac{\mu^{\prime\prime}}{2}\left(1+\mu\right)+\frac{\mu^{\prime2}}{4}\right)\psi+\phi\frac{1}{4}V\left(x,t\right)\psi\right]dx+\\
 & \int\left[-\frac{\hbar^{2}}{8m}\psi^{*}\left(\left(1+\mu^{*}\right)^{2}\partial_{x}^{2}+2\mu^{*\prime}\left(1+\mu^{*}\right)\partial_{x}+\frac{\mu^{*\prime\prime}}{2}\left(1+\mu^{*}\right)+\frac{\mu^{*\prime2}}{4}\right)\phi^{*}+\frac{1}{4}\psi^{*}V^{*}\left(x,t\right)\phi^{*}\right]dx+\\
 & \int\left[-\frac{\hbar^{2}}{8m}\phi^{*}\left(\left(1+\mu^{*}\right)^{2}\partial_{x}^{2}+2\mu^{*\prime}\left(1+\mu^{*}\right)\partial_{x}+\frac{\mu^{*\prime\prime}}{2}\left(1+\mu^{*}\right)+\frac{\mu^{*\prime2}}{4}\right)\psi^{*}+\frac{1}{4}\phi^{*}V^{*}\left(x,t\right)\psi^{*}\right]dx.
\end{align*}
 Now, we supposes that the field $\psi$ is $\mathcal{PT}$-symmetric
i. e., $\psi\left(x\right)=\psi^{*}\left(-x\right)$ and $\phi=\psi$.
Hence, the energy reduces to
\begin{equation}
E=\frac{1}{2}\left(\int\psi\hat{H}\psi dx+\int\psi^{*}\hat{H}^{\dagger}\psi^{*}dx\right),\label{exp-value}
\end{equation}
 and upon considering $\left[H,\mathcal{PT}\right]=0$ the two terms
become identical such that
\begin{equation}
E=\int\psi\hat{H}\psi dx=\int\psi^{\mathcal{PT}}\hat{H}\psi dx=\left\langle \hat{H}\right\rangle ,\label{Expectation-value}
\end{equation}
in which $\left\langle \hat{H}\right\rangle $ is the expectation
value of $H$ in $\mathcal{PT}$-symmetric quantum mechanics. 

The next quantity which can be discussed is the stress-energy tensor.
In this line, first, we obtain the energy flux density corresponding
to the Cartesian coordinate which is defined by
\begin{equation}
\mathcal{S}=\dot{\psi}\frac{\partial\mathcal{L}}{\partial\psi^{\prime}}+\dot{\phi}\frac{\partial\mathcal{L}}{\partial\phi^{\prime}}+\dot{\psi^{*}}\frac{\partial\mathcal{L}}{\partial\psi^{*\prime}}+\dot{\phi^{*}}\frac{\partial\mathcal{L}}{\partial\phi^{*\prime}}.\label{energy flux}
\end{equation}
The explicit calculation reveals that

\begin{align}
\mathcal{S} & =-\frac{\hbar^{2}}{4m}\left[\left(1+\mu\right)^{2}\left(\dot{\psi}\phi^{\prime}+\dot{\phi}\psi^{\prime}\right)+\left(1+\mu\right)\frac{\mu^{\prime}}{2}\dot{\left(\psi\phi\right)}\right]+\nonumber \\
 & -\frac{\hbar^{2}}{4m}\left[\left(1+\mu^{*}\right)^{2}\left(\dot{\psi^{*}}\phi^{*\prime}+\dot{\phi^{*}}\psi^{*\prime}\right)+\left(1+\mu^{*}\right)\frac{\mu^{*\prime}}{2}\dot{\left(\psi^{*}\phi^{*}\right)}\right].\label{Energy flux}
\end{align}
Second, we obtain the momentum density which is defined by
\begin{equation}
\boldsymbol{P}=\psi^{\prime}\frac{\partial\mathcal{L}}{\partial\dot{\psi}}+\phi^{\prime}\frac{\partial\mathcal{L}}{\partial\dot{\phi}}+\psi^{*\prime}\frac{\partial\mathcal{L}}{\partial\dot{\psi^{*}}}+\phi^{*\prime}\frac{\partial\mathcal{L}}{\partial\dot{\phi^{*}}}.\label{momentum density}
\end{equation}
Our detailed calculations lead to 
\begin{equation}
\boldsymbol{P}=\frac{i\hbar}{2}\left(\psi^{\prime}\phi-\psi^{*\prime}\phi^{*}\right).\label{momentum-density}
\end{equation}
Let's remind that the momentum density implies the amount of energy
in a unit of volume passing through a surface in the unit of time.
It represents the physical spatial momenta corresponding to a field
mutually with the canonical momentum of a quantum particle. 

\section{Continuity equation}

Next, we utilize \eqref{Schrodinger-psi} and \eqref{Schrodinger-phi}
and their complex conjugates \eqref{Schrodinger-star} and \eqref{Shrodinger-star2}
to find the continuity equation. Let's multiply by $\phi$, $\phi^{*}$
, $\psi$ and $\psi^{*}$ from the left the equations \eqref{Schrodinger-psi},
\eqref{Schrodinger-star}, \eqref{Schrodinger-phi} and \eqref{Shrodinger-star2},
respectively. Then by simple addition and subtraction of the results,
we obtain 

\begin{equation}
i\hbar\partial_{t}\left(\phi\psi+\phi^{*}\psi^{*}\right)=-\frac{\hbar^{2}}{2m}\partial_{x}\left(\left(1+\mu\right)^{2}\left(\psi\phi^{\prime}-\phi\psi^{\prime}\right)+\left(1+\mu^{*}\right)^{2}\left(\phi^{*}\psi^{*\prime}-\psi^{*}\phi^{*\prime}\right)\right).\label{Continuity}
\end{equation}
This is the continuity equation provided we define 
\begin{equation}
\rho=\phi\psi+\phi^{*}\psi^{*},\label{probability density}
\end{equation}
to be the probability density and 
\begin{equation}
j=\frac{\hbar}{2im}\left(\left(1+\mu\right)^{2}\left(\phi\psi^{\prime}-\psi\phi^{\prime}\right)-\left(1+\mu^{*}\right)^{2}\left(\phi^{*}\psi^{*\prime}-\psi^{*}\phi^{*2}\right)\right),\label{current density}
\end{equation}
 to be the particle current density. Hence the continuity equation
\begin{equation}
\frac{\partial\rho\left(x,t\right)}{\partial t}+\frac{\partial j\left(x,t\right)}{\partial x}=0\label{ContinuityStandard}
\end{equation}
holds. We would like to comment that the present outcomes in \eqref{probability density}
and \eqref{current density} are significant since the conservation
of probability density is confirmed. We note that with $\psi\left(x,t\right)=\psi^{\mathcal{PT}}\left(x,t\right)=\phi\left(x,t\right)$,
the particle current density vanishes and consequently, 
\begin{equation}
\frac{d}{dt}\int dx\rho=0,\label{conservation}
\end{equation}
which is the conservation of the total probability on $x\in\mathbb{R}$
\cite{Bagchi}. 

\section{Conclusion}

We have introduced a Lagrangian density in terms of the fields $\psi$,
$\phi$, $\psi^{*}$ and $\phi^{*}$ for a quantum particle with a
generalized momentum operator \cite{M.H1,M.H2}. By virtue of the
variation of the Lagrangian density, we obtained the Schrödinger equation
which describes the behavior of such quantum particles. Furthermore,
we calculated the Hamiltonian density and showed that the energy of
the system is the expectation value of the Hamiltonian operator. Finally,
we constructed the energy flux, the momentum density, and the probability
density as well as particle current density which the last two satisfy
the standard continuity equation.

\end{document}